\documentclass[aps,twocolumn,superscriptaddress, longbibliography]{revtex4-1}
\usepackage[english]{babel}
\usepackage{graphicx}
\usepackage{stmaryrd}
\usepackage{amsmath, amsfonts}
\usepackage{mathrsfs}
\usepackage{braket}
\usepackage{color}
\usepackage{xspace}
\usepackage{amscd}
\usepackage{latexsym}
\usepackage{diagbox}
\usepackage{bm}
\definecolor{lightblue}{rgb}{0.13, 0.26, 0.99}

\usepackage[
colorlinks=true,
urlcolor=blue,
citecolor=blue,
linkcolor=blue,
hyperfootnotes=false]{hyperref}

\allowdisplaybreaks

\begin{document}

\title{Translation constraints on quantum phases with twisted boundary conditions}
\author{Shunsuke C. Furuya}
\affiliation{Condensed Matter Theory Laboratory, RIKEN, Wako, Saitama 351-0198, Japan}
\author{Yusuke Horinouchi}
\affiliation{RIKEN Center for Emergent Matter Science (CEMS), Wako, Saitama 351-0198, Japan
}
\date{\today}
\begin{abstract}
Bulk properties of quantum phases should be independent of a specific choice of boundary conditions as long as the boundary respects the symmetries.
Based on this physically reasonable requirement, we discuss the Lieb-Schultz-Mattis-type ingappability in two-dimensional quantum magnets under a boundary condition that makes evident a quantum anomaly underlying the lattice system.
In particular, we direct our attention to those on the checkerboard lattice which are closely related to frustrated quantum magnets on the square lattice and on the Shastry-Sutherland lattice.
Our discussion is focused on the adiabatic U(1) flux insertion through a closed path in a boundary condition twisted by a spatial rotation and a reflection.
Two-dimensional systems in this boundary condition are effectively put on a nonorientable space, namely the Klein bottle.
We show that the translation symmetry on the Klein-bottle space excludes the possibility of the unique and gapped ground state.
Taking advantage of the flux insertion argument, we also discuss the ground-state degeneracy on magnetization plateaus of the Heisenberg antiferromagnet on the checkerboard lattice.
\end{abstract}
\maketitle

\section{introduction}\label{sec:intro}

Quantum critical points provide a good starting point toward understanding phases of quantum many-body systems.
In the language of quantum field theory, which effectively describes low-energy physics of quantum many-body systems, a quantum critical point corresponds to a fixed point of the renormalization group. One will thus know possible quantum phases neighboring a quantum critical point by listing possible relevant perturbations to a corresponding fixed point.

In quantum many-body systems, when listing possible quantum phases, the Lieb-Schultz-Mattis (LSM) theorem imposes a strong constraint taht excludes the possibility of a unique and gapped ground state under a certain condition~\cite{lsm, affleck_lsm, oshikawa_lsm, hastings_lsm}. 
Originally, LSM proved the absence of the unique and gapped ground state in the spin-$1/2$ Heisenberg antiferromagnetic chain~\cite{lsm}.
Later, the LSM theorem was extended to various systems on higher-dimensional lattices~\cite{affleck_lsm, oshikawa_lsm, hastings_lsm, Watanabe_lsm_pnas, po_lattice_homotopy, metlitski_lsm}.
In quantum field theories, one can make a claim corresponding to the LSM theorem on the basis of quantum anomalies such as the 't Hooft anomaly~\cite{kapustin_anomaly, cho_lsm, tanizaki_lsm, yao_lsm_anomaly}.
When a quantum field theory under a given symmetry has an anomaly, a corresponding quantum critical point is \emph{not} driven into the quantum disordered phase as long as the symmetry is maintained.

The anomaly manifests itself as an obstacle when defining the quantum field theory as an effective description of the bulk phase of quantum many-body systems. For example, the anomaly leads to an unphysical dependence on boundary conditions.
It is widely believed that bulk properties, such as the existence of the gap and the ground-state degeneracy due to the spontaneous symmetry breaking, should be independent of a specific choice of boundary conditions unless the boundary breaks the symmetry.
One of the authors in Ref.~\cite{furuya_wzw} discussed the violation of the modular invariance as the obstacle in (1+1)-dimensional systems.
The modular invariance signifies the fact that the two-dimensional conformal field theory in the bulk phase is unfettered by symmetric modifications of the boundary conditions.
Its violation is indeed unphysical.
The anomaly as the violation of the modular invariance explains consistently the LSM-type ingappability of excluding the possibility of the unique and gapped ground state in $1+1$ dimensions~\cite{furuya_wzw}.

\begin{figure}[b!]
    \centering
    \includegraphics[viewport = 0 0 1500 700, width=\linewidth]{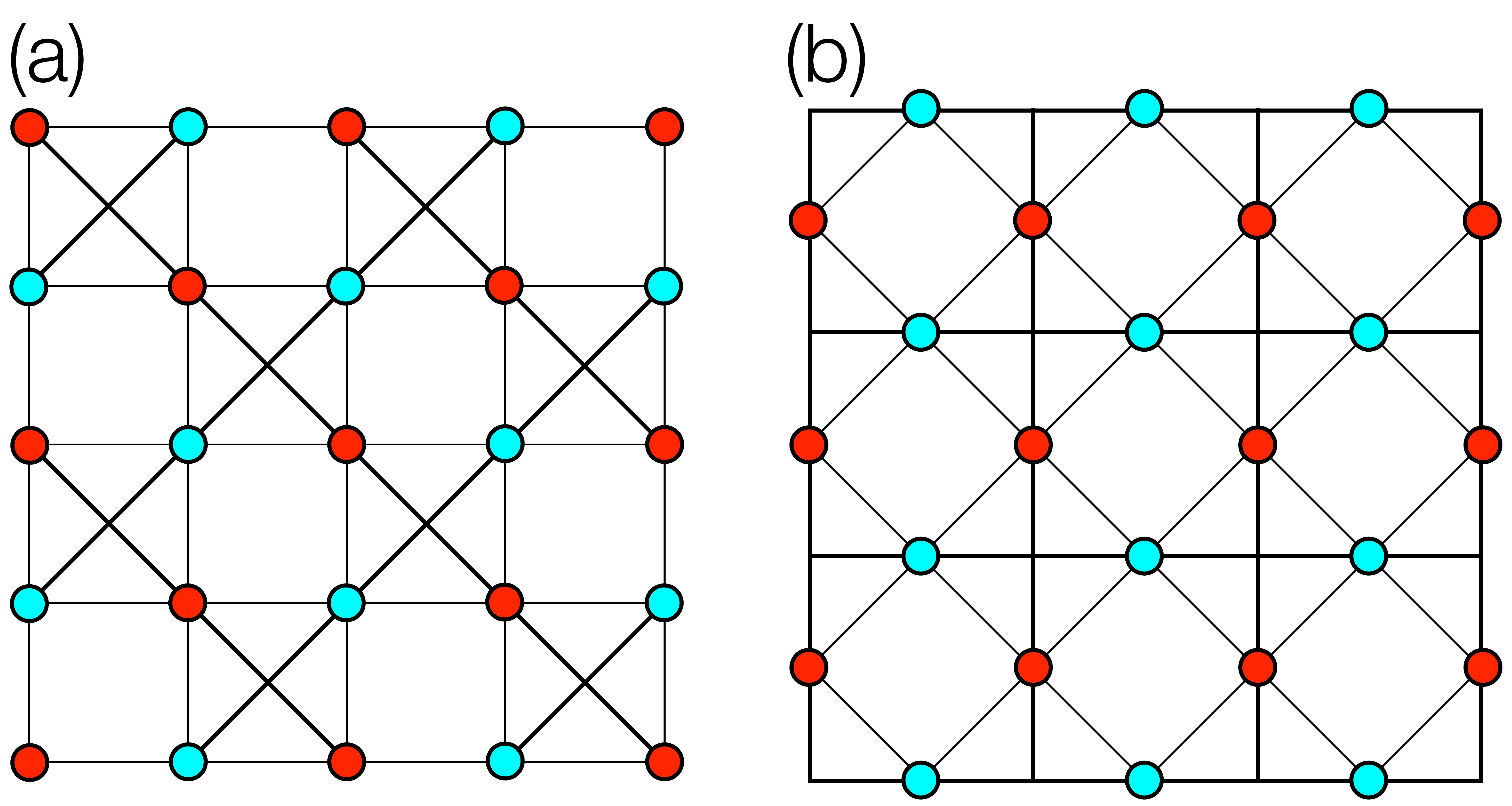}
    \caption{The checkerboard lattices. One can see the checkerboard lattice as (a) a square lattice with diagonal bonds on every other square plaquette and also as (b) a crossed chain model.}
    \label{fig:cb}
\end{figure}

It will also be interesting to extend the argument of Ref.~\cite{furuya_wzw} to higher-dimensional systems in order to foster a better understanding of relations between the anomaly and the boundary condition.
Yao and Oshikawa reported quite recently a paper that follows this line~\cite{yao_lsm_boundary}, where they adapt the flux insertion argument~\cite{laughlin_flux, oshikawa_lsm} in a ``tilted'' boundary condition  instead of the periodic one.
The tilt of the boundary condition clarifies the existence of the anomaly of, for example, the $S=1/2$ Heisenberg antiferromagnet on the $d$-dimensional hypercubic lattice ($d=2,3,\cdots$).
An appropriate choice of the boundary condition turned out to make the anomaly manifest.

It then came to our attention that there is a frustrated quantum magnets whose anomaly, though it is certainly present, is out of sight in the periodic boundary condition or in the tilted boundary condition.
It is a spin-$1/2$ Heisenberg antiferromagnet on the checkerboard lattice~\cite{starykh_checkerboard}.
One can demonstrate the presence of the anomaly of the checkerboard in several ways, for instance, by applying the lattice homotopy argument~\cite{po_lattice_homotopy}.
Nevertheless, as we show later, the flux insertion argument fails to detect the anomaly in those boundary conditions.
It is interesting by itself to construct a boundary condition that 
incarnates the anomaly on the checkerboard.
In addition, such an argument is attracting in its potential application to magnetization plateaus~\cite{oya, tanaka_geometricalphase}.
In fact, it was recently shown that the checkerboard Heisenberg antiferromagnet hosts numerous magnetization plateaus~\cite{morita_checkerboard_plateau, capponi_checkerboard_plateau}.

In this paper, we introduce translationally invariant boundary conditions that are twisted by a spatial rotation and a reflection.
In that context, we discuss the LSM-type ingappability on the checkerboard lattice as a continuation of the work in Refs.~\cite{furuya_wzw, yao_lsm_boundary}.
First, we show that the twisted boundary condition indeed enables us to detect the anomaly of the checkerboard through the flux insertion process in the presence of time-reversal symmetry.
Next, removing the time-reversal symmetry, we discuss the ground-state degeneracy on magnetization plateaus of the checkerboard.

\section{Conjecture}\label{sec:conj}

Here, we clarify the physical meaning of a conjecture made in this paper about relations between the LSM-type anomaly and boundary conditions.
In Sec.~\ref{sec:intro}, we mentioned the modular invariance (noninvariance) of the two-dimensional conformal field theory as a manifestation of the absence (presence) of the LSM-type anomaly~\cite{furuya_wzw}. 
This result naturally motivates us to make the following conjecture:
\emph{If the unique gapped ground state is allowed under a symmetry in the periodic boundary condition, the system is free from the anomaly in any other symmetric boundary conditions}.
Therefore, in order to show the ingappability, we just need to find a certain symmetric boundary condition under which the LSM-type anomaly is clearly present.
This is what we do in the remainder of the paper.
The above conjecture is also supported by the fact that the topological field theoretical classification of the anomaly is independent of boundary conditions.

\section{Checkerboard}\label{sec:cb}

\subsection{$S=1/2$ Heisenberg antiferromagnet}

Let us start with a simple spin-$1/2$ Heisenberg antiferromagnet on the checkerboard lattice [Fig.~\ref{fig:cb}~(a)].
We may regard them also as a crossed chain model [Fig.~\ref{fig:cb}~(b)]~\cite{starykh_checkerboard}.
It has the following Hamiltonian,
\begin{align}
    \mathcal H_{\rm CBH}
    &= \mathcal H_h + \mathcal H_v + \mathcal H_c,
    \label{H_cb}
\end{align}
where $\mathcal H_h$ and $\mathcal H_v$ denote the Heisenberg exchange interactions on the horizontal chains and the vertical chains, respectively, and $\mathcal H_c$ denotes frustrated interchain interactions:
\begin{align}
    \mathcal H_h &= J \sum_{n=1}^{L_h}\sum_{m=1}^{L_v} \bm S_h(n-\tfrac 12, m) \cdot \bm S_h(n+\tfrac 12, m),
    \label{H_h} \\
    \mathcal H_v
    &=J \sum_{n=1}^{L_h} \sum_{m=1}^{L_v} \bm S_v(n, m-\tfrac 12) \cdot \bm S_v(n,m+\tfrac 12),
    \label{H_v} \\
    \mathcal H_c
    &=  J_\times \sum_{n=1}^{L_h}\sum_{m=1}^{L_v} \{\bm S_h(n-\tfrac 12, m) +\bm S_h (n+\tfrac 12, m)\}
    \notag \\
    &\qquad
     \cdot \{ \bm S_v(n,m-\tfrac 12) + \bm S_v(n,m+\tfrac 12) \}.
    \label{H_c}
\end{align}

$\bm S_h(x,y)$ and $\bm S_v(x,y)$ are $S=1/2$ spins at a site $(x,y)$ on a horizontal chain and on a vertical chain, respectively.
$L_h$ and $L_v$ are the lengths of the system along the horizontal and the vertical axis of Fig.~\ref{fig:cb}~(b), respectively, in the unit of the unity lattice spacing.
The exchange couplings $J$ and $J_\times$ are both positive and thus the model \eqref{H_cb} has only antiferromagnetic interactions.

The checkerboard Heisenberg antiferromagnet \eqref{H_cb} exhibits an interesting ground-state phase diagram~\cite{starykh_checkerboard}.
For $J_\times/J\ll 1$, the ground state
exhibits a dimerization that spontaneously breaks the translation symmetry.
For $J_\times/J\gg 1$, on the other hand, it has the spontaneous N\'eel order because the checkerboard is reduced to being a simple square lattice in the limit of $J_\times/J \to + \infty$.
For a moderate $J_{\times}/J\sim 1$, the plaquette valence-bond-solid phase is realized in between the above two phases,  as validated by the exact diagonalization~\cite{fouet_checkerboard} and the density-matrix renormalization-group~\cite{morita_checkerboard_plateau} methods.
Reference~\cite{starykh_checkerboard} proposed two possible scenarios of the ground-state phase diagram in changing $J_\times /J \in [0,\infty)$, both of which contain no quantum disordered phase of the unique and gapped ground state.
It is thus natural to guess that the $S=1/2$ checkerboard Heisenberg antiferromagnet has an anomaly that prevents the ground state from being unique and gapped.

\subsection{'t Hooft anomaly}

In fact, it is shown by the classification of three-dimensional weak symmetry-protected topological (SPT) phases that the checkerboard Heisenberg antiferromagnet has the 't Hooft anomaly in the $\mathrm{U(1)} \times \mathbb Z_2^T \times \mathbb Z^2$ symmetry.
Here, $\mathrm{U(1)}$ is the $\mathrm{U(1)}$ spin rotation symmetry, $\mathbb Z_2^T$ is the time-reversal symmetry, and $\mathbb Z^2$ is the translation symmetry in the horizontal and the vertical axes.
The field-theoretical derivation of the classification is given in Appendix~\ref{app:cobordism} and \ref{app:top}.
The 't Hooft anomaly implies that under the time-reversal symmetry, the U(1) gauge transformation will be incompatible with the translation symmetry.
This anomaly is expected to appear in the argument of the adiabatic flux insertion~\cite{oshikawa_lsm}.
However, as we mentioned, the flux insertion method under the periodic boundary condition does not show it clearly.

\subsection{Generic spin-$S$ quantum magnets}

Let us first demonstrate that the U(1)-flux insertion developed in Ref.~\cite{oshikawa_lsm} fails to capture the anomaly on the checkerboard.

\subsubsection{In the periodic boundary condition}

In what follows, we consider the spin-$S$ checkerboard Heisenberg antiferromagnet for general $S\ge 1/2$.
First, we rewrite the model \eqref{H_cb} as a bilayer system [Fig.~\ref{fig:cb}~(b)] where the red circles are located on the upper layer and the light blue ones are on the lower layer.
Let us assign a new label to the spin,
\begin{align}
    \bm S_{n,m,1} &:= \bm S_v(n, m+\tfrac 12) \\
    \bm S_{n,m,2} &:= \bm S_h(n-\tfrac 12, m).
\end{align}
$\bm S_{n,m,l}$ satisfies the periodic boundary condition:
\begin{align} 
    \left\{
    \begin{array}{rl}
        \bm S_{n+L_h,m, l} &= \bm S_{n,m,l}, \\[+10pt]
         \bm S_{n, m+L_v, l}&= \bm S_{n,m,l},
    \end{array}
    \right.
    \label{pbc}
\end{align}
for $l=1,2$.

Next, we pierce the system with the flux by replacing the $xy$ component of the exchange interactions as 
\begin{align}
    &S_{n,m, l}^+S_{n+1,m',l'}^- +\mathrm{H.c.} 
    \notag \\
    &\qquad \longrightarrow \exp(-i\tfrac{\Phi}{L_h}) S_{n,m,l}^+S_{n+1,m',l'}^- + \mathrm{H.c.}
\end{align}
Note that $S_{n,m,l}^\pm := S_{n,m,l}^x \pm i S_{n,m,l}^y$.
We increase $\Phi$ adiabatically from zero to $2\pi$, i.e., the unit amount.
The unit flux is erased by a U(1) large gauge transformation generated by
\begin{align}
    U_{\rm P} &= \exp\biggl(i\frac{2\pi}{L_h} \sum_{n=1}^{L_h}\sum_{m=1}^{L_v} n\sum_{l=1,2}  (S-S_{n,m,l}^z) \biggr).
    \label{u_pbc}
\end{align}
Let $\ket{\Psi_0}$ be a ground state of the checkerboard Heisenberg antiferromagnet without the flux.
The adiabatic insertion of the flux metamorphoses $\ket{\Psi_0}$ eventually into $\ket{\Psi'_0}$, which is a ground state of the checkerboard Heisenberg antiferromagnet with the unit flux.
$U_{\rm P} \ket{\Psi'_0}$ is then a ground state of the original checkerboard Heisenberg antiferromagnet without the flux.
In general, if one can find a conserved charge $\mathcal O$ that does not commute with $U_{\rm P}$, the system cannot have the unique and gapped ground state. Namely, $\ket{\Psi_0}$ and $U_P\ket{\Psi'_0}$ are orthogonal to each other since they are distinguished by the eigenvalues of $\mathcal O$.
In the LSM paper~\cite{lsm} and Ref.~\cite{oshikawa_lsm}, they chose $\mathcal O$ as the one-site translation along the direction on which the periodic boundary condition is imposed.
This corresponds in our case to $\mathcal O= T_h$ defined by a relation $T_h \bm S_{n,m,l} T_h^{-1} = \bm S_{n+1,m,l}$, in our case.
From the relation,
\begin{align}
    T_h U_{\rm P}T_h^{-1}
    &= U_{\rm P} \exp\biggl(-i\frac{2\pi}{L_h}\sum_{n=1}^{L_h}\sum_m \sum_{l=1,2}(S - S_{n,m,l}^z) \biggr),
\end{align}
it follows that $T_h$ and $U_{\rm P}$ commute with each other as long as the ground state has the zero total magnetization.
Therefore, we cannot deduce the expected ground-state degeneracy by a flux insertion under the boundary condition of Eq.~\eqref{pbc}.

\subsubsection{In the tilted boundary condition}

Following the requirement of insensitivity of the bulk phase to boundary conditions,
we replace the boundary condition and keep track of the translation symmetry.
Let us impose the tilted boundary condition on the system~\cite{yao_lsm_boundary}.
The tilted boundary condition is defined as
\begin{align}
    \left\{
    \begin{array}{rl}
        \bm S_{n+L_h,m,l} &= \bm S_{n,m+1,l},   \\[+10pt]
        \bm S_{n,m+L_v,l} &= \bm S_{n,m,l},
    \end{array}
    \right.
    \label{tilt_bc}
\end{align}
for $l=1,2$.
Under the tilted boundary condition, we can sweep the whole checkerboard lattice by performing the one-site translation $T_h$ iteratively.
The tilted boundary condition allows us to regard the checkerboard as a one-dimensional ring on which all the $2L_hL_v$ sites are located.
We then pierce the system as the ring adiabatically with the flux until it reaches the unit amount.
The unit flux is erased by a U(1) large gauge transformation,
\begin{align}
    U_{\rm T}&= \exp\biggl(i\frac{2\pi}{L_hL_v} \sum_{r_1=1}^{L_hL_v} r_1 \sum_{l=1,2} (S-S_{n,m,l}^z) \biggr).
\end{align}
Here, $r_1=1,2, \cdots, L_hL_v-1, L_hL_v$ is a label of the site along the ring, and it is related to the two-dimensional coordinate $(n,m)$ for $n \in [1,L_h]$ and $m\in [1,L_v]$ through
\begin{align}
    r_1 &= n + (m-1)L_h
\end{align}
It immediately follows that
\begin{align}
    T_h U_{\rm T}T_h^{-1}
    &= U_T\exp\biggl(-i\frac{2\pi}{L_hL_v} \sum_{n,m} \sum_{l=1,2} (S - S_{n,m,l}^z) \biggr).
\end{align}
Again, we obtain $T_h U_{\rm T} T_h^{-1} = U_{\rm T}$ in the absence of the total magnetization
and the expected ground-state degeneracy cannot be deduced.

\section{Spatially twisted boundary conditions}\label{sec:rot}

\begin{figure}[t!]
    \centering
    \includegraphics[viewport = 0 0 2000 2000, width=\linewidth]{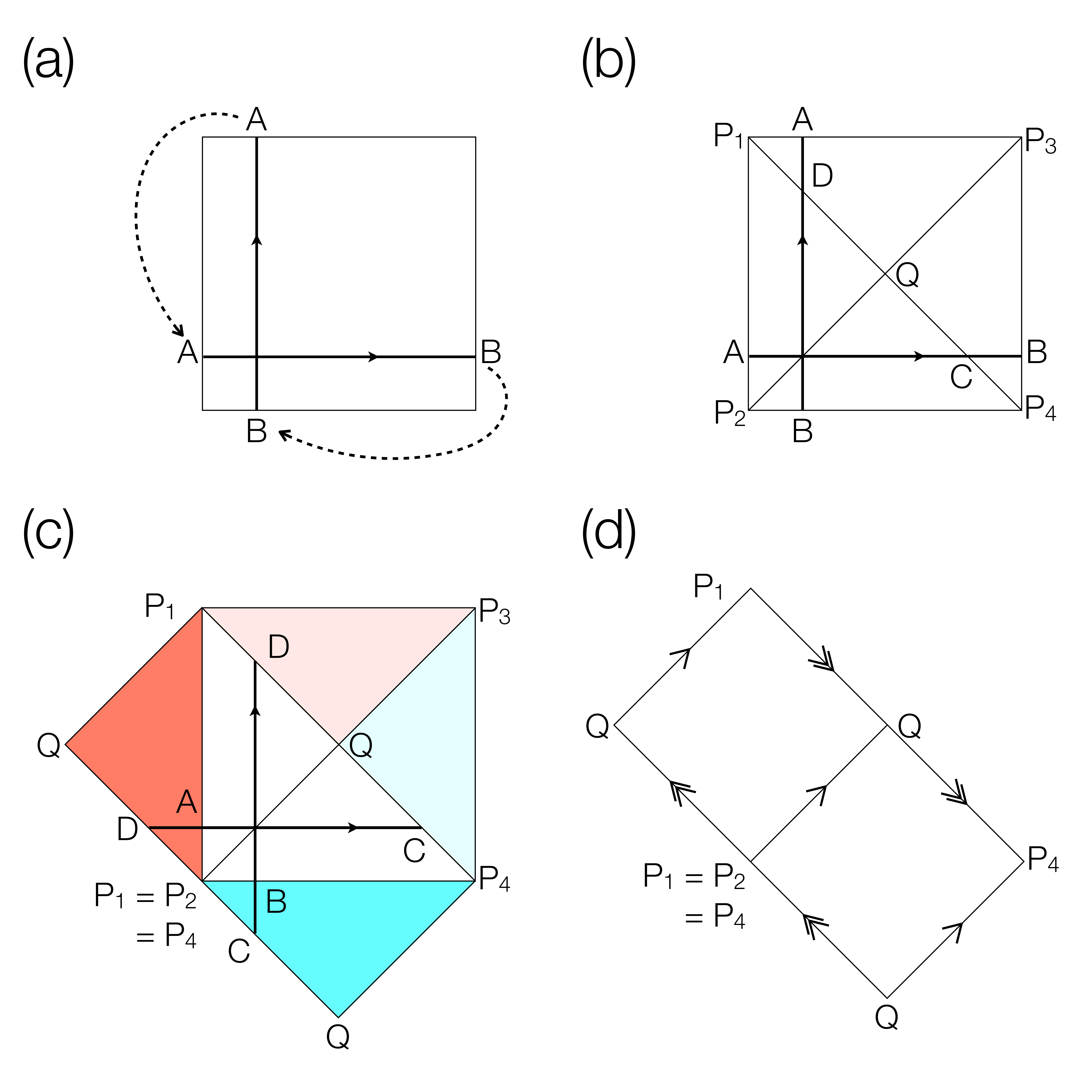}
    \caption{The Klein-bottle boundary condition on the two-dimensional plane. (a) Sites at boundaries labeled by the same symbol are identified in the Klein-bottle boundary condition. (b,c) The square is cut into four equal-area parts and recombined to the rectangle. (d) The rectangle so produced is made of a pair of M\"obius strips and is equivalent to the Klein bottle.}
    \label{fig:kb}
\end{figure}

In the previous section, we found that the periodic or the tilted boundary condition fails to make the anomaly clear in the flux insertion.
Here, in this section, we introduce another symmetric boundary condition that enables the flux insertion along with a shift of the crystal momentum.

\subsection{The Klein bottle and the tilted Klein bottle}

First, we introduce a Klein-bottle boundary condition shown in Fig.~\ref{fig:kb}~(a).
When we reach the right edge of the system, we reenter the system from the bottom edge.
This boundary condition is more precisely defined as
\begin{align}
    \left\{
    \begin{array}{rl}
        \bm S_h(n-\tfrac 12 + L_h, m) &= \bm S_v(m,n-\tfrac 12), \\[+10pt]
        \bm S_v(n,m-\tfrac 12 + L_v) &= \bm S_h(m-\tfrac 12, n).
    \end{array}
    \right.
    \label{kb}
\end{align}
Namely, this boundary condition is twisted by a spatial rotation $\bm S(x,y) \mapsto \bm S(y,-x)$ and a spatial reflection $\bm S(y,-x) \mapsto \bm S(y,x)$.
As a consequence of this geometrical operation, the boundary condition \eqref{kb} is valid when
\begin{align}
    L_h = L_v = L.
    \label{cond_sq}
\end{align}

Imposing the boundary condition \eqref{kb} on the system is equivalent to putting the system on the Klein bottle.
To see this, we divide the system into four equal-area parts [Fig.~\ref{fig:kb}~(b)] and recombine them into a rectangle as shown in Fig.~\ref{fig:kb}~(c).
The rectangular system is made of two M\"obius strips [Fig.~\ref{fig:kb}~(d)] and is equivalent to the Klein bottle.
The Klein-bottle boundary condition, as well as the tilted one, maintains the one-site translation symmetry across the seam of the system.

Next, we tilt the boundary and modify the Klein-bottle boundary condition to
\begin{align}
    \left\{
    \begin{array}{rl}
        \bm S_h(n-\tfrac 12 + L, m) &= \bm S_v(m,n-\tfrac 12), \\[+10pt]
        \bm S_v(n,m-\tfrac 12 + L) &= \bm S_h(m-\tfrac 12, n+1).
    \end{array}
    \right.
    \label{tkb}
\end{align}
Relation \eqref{tkb} consists of the twisting operation of \eqref{kb} and the tilt.
Thus, we call the boundary condition \eqref{tkb} a tilted Klien-bottle boundary condition.

The Klein-bottle boundary conditions \eqref{kb} and the tilted Klein-bottle boundary condition \eqref{tkb} are compatible with the square system \eqref{cond_sq} though the length on one side can be even or odd integers in the unit of the unity lattice spacing.
The restriction in the shape will be irrelevant in the thermodynamic limit.
However, we will discuss in Sec.~\ref{sec:plateau} that the restriction in the shape can be crucial in predicting the ground-state degeneracy of finite-size systems.

\subsection{Flux insertion}

The tilted Klein-bottle boundary condition \eqref{tkb} allows a one-dimensional sweep of the checkerboard lattice just as the tilted boundary condition \eqref{tilt_bc} does.
The only and crucial difference in these boundary conditions is the number of layers, or the number of spins in the unit cell.
In the tilted boundary condition of the previous section, the unit cell contains two sites.
In the tilted Klein-bottle boundary condition, the unit cell contains only a single site.

Let us pierce the checkerboard in the tilted Klein-bottle boundary condition with the U(1) flux and erase it by the following U(1) large gauge transformation,
\begin{align}
    U_{\rm R} &= \exp\biggl(i\frac{2\pi}{2L^2} \sum_{r'_1=1}^{2L^2} r'_1 (S- s_{r'_1}^z) \biggr),
\end{align}
where $r'_1=1,2, \cdots, 2L^2-1, 2L^2$ is the one-dimensional coordinate to specify the location of the spin $\bm s_{r'_1}$, which corresponds to $\bm S_h(n-\tfrac 12, m)$ and $\bm S_v(n,m-\tfrac 12)$ in the following manner.
\begin{align}
    \bm S_h(n-\tfrac 12,m) &= \bm s_{n+2(m-1)L}, \\
    \bm S_v(n,m-\tfrac 12) &= \bm s_{m+(2n-1)L}.
\end{align}

The tilted Klein-bottle boundary condition defines the one-dimensional path sweeping the whole checkerboard lattice.
The one-site translation along the path, which we call $T_r$, acts on $\bm s_{r'_1}$ as
\begin{align}
    T_r \bm s_{r'_1} T_r^{-1} &= \bm s_{r'_1+1}.
\end{align}
It is then obvious that $T_r$ and $U_{\rm R}$ satisfy
\begin{align}
    T_r U_{\rm R}T_r^{-1}
    &= U_{\rm R}\exp\biggl(-i\frac{2\pi}{2L^2} \sum_{n,m=1}^{2L^2} (S-s_{r'_1}^z )\biggr).
\end{align}
In the absence of the total magnetization, we obtain
\begin{align}
    T_r U_{\rm R} T_r^{-1} &=  U_{\rm R} \exp(- 2\pi Si).
\end{align}
Therefore, for any half-odd-integer $S$, the translation $T_r$ and the U(1) large gauge transformation $U_{\rm R}$ do not commute with each other.
We reach the conclusion that the spin-$S$ Heisenberg antiferromagnet on the checkerboard lattice cannot have a unique and gapped ground state when $S\in \mathbb Z+1/2$.
The anomaly related to the LSM-type ingappability, which we call the LSM-type anomaly, is $\mathbb Z_2$ in our case.

\subsection{Symmetric and asymmetric modifications of the model}

We may add various interactions to the spin-$S$ checkerboard Heisenberg antiferromagnet without affecting the anomaly as long as those interactions maintain the symmetries.

\subsubsection{Symmetric modifications}

One can modify the checkerboard Heisenberg model to a frustrated  square-lattice Heisenberg model by adding an interaction
\begin{align}
    &J \sum_{n,m} \{ \bm S_h(n-\tfrac 12,m) \cdot \bm S_h(n-\tfrac 12,m+1)
    \notag \\
    &\qquad + \bm S_v(n,m-\tfrac 12) \cdot \bm S_v(n+1,m-\tfrac 12)
    \},
    \label{sym_int}
\end{align}
to the Hamiltonian \eqref{H_cb}.
The resultant model is the so-called $J_1$-$J_2$ model on the square lattice where the nearest-neighbor exchange coupling is $J_1=J_\times$ and the next-nearest-neighbor one is $J_2=J$.

The ground-state phase diagram of the spin-$1/2$ $J_1$-$J_2$ model has been numerically discussed for many years.
Obviously, the ground state is in the N\'eel ordered phase for $0\le J_2/J_1\ll 1.$
When the ratio $J_2/J_1$ is increased, the system undergoes a quantum phase transition and enters into a phase different from the N\'eel one.
The nature of this phase has long been discussed and is still controversial.
In fact, there are many proposals for that phase such as the gapped spin-liquid phase~\cite{gong_square_gappedSL_2012, mezzacapo_square_gappedSL_2012}, the gapless spin-liquid phase~\cite{capriotti_square_gaplessSL_2001, Hu_square_gaplessSL_2013, wang_square_gaplessSL_2013, gong_square_gaplessSL_2014, wang_square_gaplessSL_2018} and a columner valence-bond-crystal phase~\cite{haghsenas_square_cvbs_2018}.

From the viewpoint of flux insertion, the $J_1$-$J_2$ frustrated square-lattice Heisenberg model is incapable of having a unique and gapped ground state because the interaction \eqref{sym_int} is $\mathrm{U(1)}\times \mathbb Z_2^T \times \mathbb Z^2$ symmetric.
The spin-$1/2$ $J_1$-$J_2$ square-lattice Heisenberg antiferromagnet has either a gapless ground state or gapped degenerate ground states.
This conclusion on the $J_1$-$J_2$ model can also be obtained in the tilted boundary condition.

\subsubsection{Asymmetric modifications}

The anomaly exists in the one-site translation symmetry of $T_r$ and the U(1) large gauge symmetry,
in the presence of time-reversal symmetry.
Therefore, breaking either the translation or the U(1) symmetry permits the unique and gapped ground state.
As we show soon below, the breakdown of the translation is interesting in its relation to the well-known Shastry-Sutherland lattice~\cite{shastry-sutherland}.

The one-site translation in a direction is easily broken by the introduction of a bond alternation in that direction.
Let us demonstrate that the bond alternation renders the ground state unique and gapped by taking as an example the $S=1/2$ checkerboard Heisenberg antiferromagnet for $0\le J_\times/J \ll 1$.
For $J_\times/J=0$, where the system is reduced to be a composition of decoupled spin chains, 
the statement is obviously true.
The bond alternation in an each spin chain opens the gap without breaking any symmetry spontaneously.
This is clearly shown in the bosonization scheme~\cite{giamarchi_book}.

Let us examine $J_\times/J$.
Keeping the most relevant interaction allowed by the $\mathrm{U(1)}\times \mathbb Z_2^T\times \mathbb Z^2$ symmetry,
we can approximate the interchain interaction \eqref{H_c}  effectively as~\cite{starykh_checkerboard}
\begin{align}
    \mathcal H_c &\approx  \sum_{n,m} g_{n,m} (-1)^{n+m} \epsilon_{h,m}(na) \epsilon_{v,n}(ma),
    \label{vx}
\end{align}
where  $\epsilon_{h,m}(na) := (-1)^n \bm S_h(n-\tfrac 12, m) \cdot \bm S_h(n+\tfrac 12, m)$ and $\epsilon_{v,n}(ma) := (-1)^m \bm S_v(n, m-\tfrac 12) \cdot \bm S_v(n,m+\tfrac 12)$ are the dimerization operator on the $m$-th horizontal chain and that on the $n$-th vertical chain, respectively, and $g_{n,m} \in \mathbb R$.
We can assume $g_{n,m}>0$ without loss of generality.
The relevant interaction \eqref{vx} pins $(\epsilon_{h,m}(na), \epsilon_{v,n}(ma))$ to either $\braket{\epsilon_{h,m}(na)} > 0 > \braket{\epsilon_{v,n}(ma)}$ or $\braket{\epsilon_{h,m}(na)} < 0 < \braket{\epsilon_{v,n}(ma)}$.
Thus, the relevant interaction \eqref{vx} drives the model into a spontaneously dimerized phase, a crossed-dimer phase~\cite{starykh_checkerboard}, 
resulting in the double degeneracy of the ground state with a finite excitation gap.

Now, we break the translation symmetry, say, in the horizontal axis.
Then it is permissible to add to $V_\times$ an interaction,
\begin{align}
    g_h\sum_{m} \int dx \, (-1)^m \epsilon_{h,m}(x),
    \label{vh}
\end{align}
where $x=na$.
Clearly, the interaction \eqref{vh} lifts the aforementioned double degeneracy and renders the ground state trivially dimerized, that is, unique and gapped.
The same occurs when breaking the translation symmetry in the vertical axis.

\begin{figure}[t!]
    \centering
    \includegraphics[viewport = 0 0 1500 700, width=\linewidth]{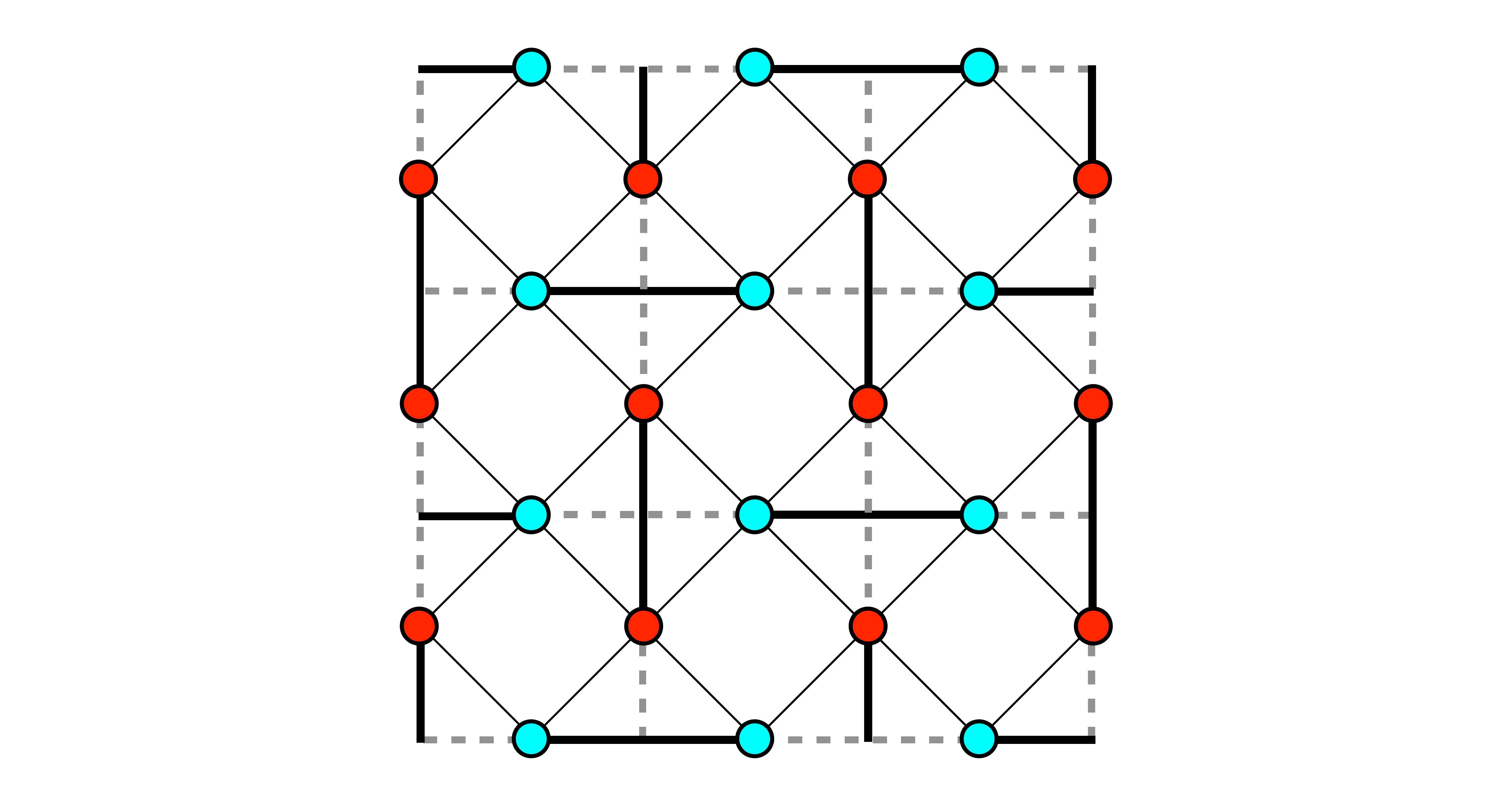}
    \caption{The bond-alternating Heisenberg antiferromagnet on the checkerboard lattice. 
    The thick bonds represent the stronger exchange interaction of $J(1+\delta)$, and the dashed bonds represent the weaker exchange interaction of $J(1-\delta)$ for $0\le \delta \le 1$.
    When all the dashed bonds are removed (i.e. when $\delta=1$), the lattice is reduced to the Shastry-Sutherland one.
    }
    \label{fig:cb2ss}
\end{figure}

Note that the bond alternation bridges the checkerboard lattice and the Shastry-Sutherland lattice.
Let us add the following bond alternation to the checkerboard Heisenberg antiferromagnet \eqref{H_cb}:
\begin{align}
    \delta \mathcal H' &= -J\delta \sum_{n,m} (-1)^{n+m}\{  \bm S_h(n-\tfrac 12,m) \cdot \bm S_h(n+\tfrac 12,m)
    \notag \\
    &\qquad - \bm S_v(n,m-\tfrac 12) \cdot \bm S_v(n,m+\tfrac 12)
    \}.
\end{align}
We depicted the model with the Hamiltonian 
\begin{align}
    \mathcal H_\delta &= \mathcal H_{\rm CBH} + \delta \mathcal H',
    \label{H_delta}
\end{align}
in Fig.~\ref{fig:cb2ss}.
Let us increase $\delta$ from $0$ to $1$.
The model $\mathcal H_{\delta=0}$ is the original Heisenberg antiferromagnet on the checkerboard lattice.
On the other hand, the model $\mathcal H_{\delta=1}$ is the Heisenberg antiferromagnet on the Shastry-Sutherland lattice.
As it is well known~\cite{shastry-sutherland}, the $S=1/2$ Heisenberg antiferromagnet on the Shastry-Sutherland lattice has a unique and gapped ground state where all the thick bonds in Fig.~\ref{fig:cb2ss} are paved with the singlet-dimer states.
It is consistent with our anomaly argument that the interaction \eqref{H_delta} breaks the translation symmetry for $\delta\not=0$ and thus removes the anomaly.

\section{Anomaly as a 1D system}\label{sec:1d}

Our discussion on the flux insertion is consistent with the topological field-theoretical classification of the LSM-type anomaly explained in appendices~\ref{app:cobordism} and \ref{app:top}.
The latter discussion is focused on the relation of the LSM-type anomaly in ($2+1$)-dimensional systems to the surface anomaly of ($3+1$)-dimensional systems in a weak SPT phase.

On the other hand, the tilted Klein-bottle boundary condition defines a one-dimensional closed path along which all the sites are swept once.
The tilted Klein-bottle boundary condition thus allows us to view the ($2+1$)-dimensional system on the checkerboard as a ($1+1$)-dimensional system with the periodic boundary condition.
Our conclusion should be independent of such a difference in viewpoints of the system.
However, this independence is \textit{a priori} nontrivial in terms of the topological field theory.

Let us briefly show that even when we regard the system as ($1+1$)-dimensional, we obtain the same LSM-type anomaly.
The proposition F.7 of Ref.~\cite{Xi17} leads to
\begin{align}
    h^D\bigl(B(G\times \mathbb Z)\bigr)
    &= h^{D-1}(BG) \oplus h^D(BG),
\end{align}
where $h^D(BG)$ refers to a generalized cohomology theory that classifies SPT phases protected by $G$ symmetry in $D$-dimensional spacetime.
The term $h^D(BG)$ on the right hand side is related to the surface anomaly of SPT phases and not of weak SPT phases.
Thus, whem we focus on the sector that is relevant in the weak SPT phase, we obtain the relation
\begin{align}
    \left. h^D(B(G\times \mathbb Z)) \right|_{\rm weak} &= h^{D-1}(BG).
    \label{eq:red}
\end{align}
The relation \eqref{eq:red} indicates that the anomaly resulting from the translational symmetry in $D$-dimensional spacetime can be detected as an anomaly in ($D-1$)-dimensional spacetime.
If we take $G$ as $\mathrm{U(1)}\times \mathbb Z$ and $h^D$ as $D\Omega^{D+1}_O$, we obtain the expected independence of the viewpoint of our system.
In fact, the $\mathbb Z_2$ anomaly discussed in Sec.~\ref{sec:rot} is related to the $\mathbb Z_2$ group as a subgroup of $D\Omega^3_O(B\mathrm{U}(1))$, which is also a subgroup of both of $D\Omega^5_O(B(\mathrm{U(1)}\times \mathbb Z^2))$ and $D\Omega^4_O(B(\mathrm{U(1)}\times \mathbb Z))$.
The former classifies the LSM-type anomaly in time-reversal symmetry, $U(1)$ symmetry, and $\mathbb Z^2$ translation symmetry in $2+1$ dimensions.
The latter classifies the LSM-type anomaly in the time-reversal symmetry, U(1) symmetry, and one-dimensional $\mathbb Z$ translation symmetry in $1+1$ dimensions.

\section{Magnetization plateaus}\label{sec:plateau}

In Secs.~\ref{sec:cb}, \ref{sec:rot} and \ref{sec:1d}, we discussed the LSM-type ingappability in the presence of the time-reversal symmetry.
Here, in this section, we break the time reversal by imposing the magnetic field on the $S=1/2$ checkerboard Heisenberg model in order not to interfere with the $\mathrm U(1)$ spin-rotation symmetry.

The magnetization curve of the $S=1/2$ checkerboard Heisenberg model was discussed in Refs.~\cite{richter_checkerboard, morita_checkerboard_plateau, capponi_checkerboard_plateau}, where numerous magnetization plateaus were found.
The $S=1/2$ checkerboard Heisenberg antiferromagnet hosts magnetization plateaus at $M/M_{\rm sat} = 1/4$ , $3/8$, $1/2$, and $3/4$,
where $M$ and $M_{\rm sat}$ are the total magnetization and its saturated value.

In the presence of the total magnetization $M>0$, the operator $U_{\rm R}$ satisfies
\begin{align}
    T_r U_{\rm R}T_r^{-1}
    &= U_{\rm R} \exp[-2\pi i (S-m)],
    \label{tut_plateau_tkb}
\end{align}
where $m=M/2L^2$ is the magnetzation density.
Generically, when $S-m= p/q$ with positive integers $p$ and $q$ which are coprime to each other, the relation \eqref{tut_plateau_tkb} claims that the ground state is at least $q$-fold degenerate~\cite{oshikawa_lsm}.
The degenerate ground states are given by $\ket{\Psi_0}$ and $U^s\ket{\Psi'_0}$ for $s=1,2, \cdots, q-1$.

On the magnetization plateau, for example, at the $3/8$ plateau, there are least 16-fold degenerate ground states because $S-m = 5/16$.
This prediction of the ground-state degeneracy on the plateau is consistent with the numerical observation~\cite{morita_checkerboard_plateau}.
However, in general one must be careful about the geometrical shape of the system when comparing the flux insertion argument \eqref{tut_plateau_tkb} with numerical results.
Numerical calculations are often performed on a finite-size cluster.
Once the shape of the cluster is fixed, the ground-state degeneracy is expected to be independent of the choice of the boundary condition.
Still, the ground-state degeneracy is in general dependent on the shape of the cluster.
For example, the relation \eqref{tut_plateau_tkb} predicts the at least eightfold degeneracy of the ground state on the $1/4$ plateau while only the fourfold degeneracy was numerically observed~\cite{capponi_checkerboard_plateau}.
Actually, the finite-size clusters used in Refs.~\cite{morita_checkerboard_plateau} and \cite{capponi_checkerboard_plateau} are incompatible with the tilted Klein-bottle boundary conditions because they are not the square defined in Eq.~\eqref{cond_sq}.
Instead, those clusters are compatible with the tilted boundary condition \eqref{tilt_bc}.
If we employ the tilted boundary condition, we obtain
\begin{align}
    T_h U_{\rm T}T_h^{-1} &= U_{\rm T} \exp[-4\pi i (S-m)],
    \label{tut_plateau_tilt}
\end{align}
where $m=M/2L_hL_v$.
Then, we conclude on the basis of Eq.~\eqref{tut_plateau_tilt} that the ground state on the $1/4$ magnetization plateau is at least 4-fold degenerate in the tilted boundary condition because $2(S-m)=3/4$.
This prediction is consistent with the numerical finding~\cite{capponi_checkerboard_plateau}.
It will be interesting to check the ground-state degeneracy numerically on the magnetization for a square-shape cluster checkerboard lattice [Fig.~\ref{fig:cb}~(b)] compatible with the tilted Klein-bottle boundary condition \eqref{tkb}.
However, this problem is beyond the scope of this paper and we leave it for future works.

\section{Summary}\label{sec:summary}

We discussed the LSM-type ingappability in two-dimensional frustrated quantum antiferromagnets.
Our discussion was focused on the anomaly between U(1) spin-rotation symmetry and translation symmetry and the physically reasonable conjecture explained in Sec.~\ref{sec:conj}.
First, we considered time-reversal symmetric cases.
In the presence of time-reversal symmetry, LSM-type ingappability is expected in the generic argument based on the 't Hooft anomaly and the surface anomaly of the weak SPT phase. 
Nevertheless, the well-known flux insertion argument turned out not to  demonstrate the anomaly explicitly in the periodic~\cite{oshikawa_lsm} or the tilted~\cite{yao_lsm_boundary} boundary conditions.
Instead of them, we imposed another boundary condition on the two-dimensional system, which is connected to the spatial rotation and the spatial reflection.
In the twisted boundary condition, which we call the tilted Klein-bottle boundary condition, the flux insertion successfully demonstrated the intrinsic LSM-type anomaly between U(1) and translation symmetries.
In particular, we showed that quantum magnets with $\mathrm{U(1)}\times \mathbb Z_2^T \times \mathbb Z^2$ symmetry on the checkerboard lattice cannot have a unique and gapped ground state.
If the ground state is gapped, it is at least doubly degenerate as a consequence of the $\mathbb Z_2$ LSM-type anomaly.

Next, we discussed the magnetization plateau in the absence of time-reversal symmetry.
Taking advantage of the flux insertion argument, we discussed the ground-state degeneracy on magnetization plateaus of the $S=1/2$ checkerboard Heisenberg antiferromagnet.
While we explained the numerically found degeneracy on some plateaus at $M/M_{\rm sat}=0, 1/2, 3/8$~\cite{morita_checkerboard_plateau, capponi_checkerboard_plateau}, we could not on the plateaus at  $M/M_{\rm sat}=1/4$ and $3/4$.
We concluded in Sec.~\ref{sec:plateau} that this disagreement originates from the shape of the finite-size cluster.
Though the bulk properties should be independent of the choice of the boundary condition, it can be dependent on the geometric shape of the system.
We emphasize that the 16-fold degeneracy of the $3/8$ plateau is explained by the flux insertion argument in the tilted Klein-bottle boundary condition but not in the periodic or the tilted boundary condition.

The tilted Klein-bottle boundary condition is applicable to any two-dimensional quantum many-body systems on square-like lattices.
Just as Ref.~\cite{yao_lsm_boundary} did in the tilted boundary condition, we can extend the tilted Klein-bottle boundary condition to higher dimensions straightforwardly, though such a higher-dimensional tilted ``Klein-bottle'' boundary condition should be unrelated to the Klein bottle directly.
In particular, it will be an interesting problem to apply the tilted ``Klein-bottle'' boundary condition to three-dimensional systems whose anomaly is less understood yet, but this reamins an open problem.

\section*{Acknowledgments}

The authors thank Yohei Fuji, Akira Furusaki, Masaki Oshikawa, Tokuro Shimokawa, Ken Shiozaki, and Yasuhiro Tada for useful discussions.

\appendix

\section{Classification of the $\mathrm{U(1)}\times \mathbb Z_2^T\times \mathbb Z^2$ anomaly}\label{app:cobordism}

In appendices~\ref{app:cobordism} and \ref{app:top}, we give a field theoretical interpretation of the LSM-type ingappability presented in the main text. 
Our discussion here is based on Refs.~\cite{CZBVB16,JBX18,TE18}, where the correspondence between weak SPT phases and LSM-type ingappability is discussed.
In the main text, we summarized the LSM-type ingappability as follows.
For a time-reversal symmetric ground state, U(1)-gauge transformation produces a nonzero momentum, that is, the gauge transformation is not compatible with translational symmetry. 
It is then natural to consider that the LSM-type anomaly can be identified with the 't Hooft anomaly of U(1)$\times\mathbb{Z}_2^T\times\mathbb Z^2$ symmetry, where $\mathbb{Z}_2^T$ and $\mathbb Z^2$ represent time-reversal and the lattice-translation symmetries, respectively.

Based on this viewpoint, we identify the 't Hooft anomaly that describes the ingappability, and we give a physically reasonable interpretation of the anomaly.
For this purpose, we first give a cobordism classification of the 't Hooft anomaly of U(1)$\times\mathbb{Z}_2^T\times\mathbb{Z}^2$ symmetry.  
In the classification, the observed LSM-type ingappability is attributed to the surface anomaly of the Haldane phase of the spin-1 Heisenberg antiferromagnetic chain.

Since every SPT phase supports an anomalous boundary, it is widely believed that the classification of the 't Hooft anomaly in the $D$-dimensional spacetime is given by that of SPT phases in the $(D+1)$-dimensional spacetime. 
Based on this physically sound assumption,  here we classify SPT phases protected by U(1)$\times\mathbb{Z}_2^T\times\mathbb{Z}^2$ symmetry in order to classify the anomaly in the lower dimension. 
According to Ref.~\cite{FH16}, bosonic SPT phases with $G$ symmetry and time-reversal symmetry in $D+1$ dimensions are classified by the Anderson dual $D\Omega^{D+2}_{O}(BG)$~\cite{An69,Yo75} of the unoriented bordism group on the classifying space $BG$~\footnote{It is believed that the SPT phases are classified by a generalized cohomology theory. A physical motivation of the generalized cohomology hypothesis is originally given by Kitaev~\cite{Ki11, Ki13, Ki15}. We also refer to Ref.~\cite{Xi17}.}.
According to the Proposition F.7 in Ref.~\cite{Xi17}, we find
\begin{align}
    &D\Omega^d_O(B(\mathrm{U(1)}\times \mathbb Z^2))
    \notag \\
    &= D\Omega^{d-2}_O(B\mathrm{U(1)})\oplus \bigl[D\Omega^{d-1}_O(B\mathrm{U(1)})\bigr]^2\oplus D\Omega^d_O(B\mathrm{U(1)}),
    \label{eq:dec}
\end{align}
where $D\Omega^{d}_O(B\mathrm{U(1)})$ is obtained from the universal property of the Anderson dual \cite{FH16} and the bordism groups~\cite{Kap14} as shown in table~\ref{tbl:MO}.
The cobordism group $D\Omega^d_O(B(\mathrm{U(1)}\times \mathbb Z^2))$ is immediately obtained from Eq.~\eqref{eq:dec} and table~\ref{tbl:MO}, which is shown in table.~\ref{tbl:MO2}.

\begin{table}[t!]
    \centering
    \begin{tabular}{c*{6}{c}}\hline \hline
         $d$ & \makebox[2em]{0} & \makebox[2em]{1} & \makebox[2em]{2}  & \makebox[2em]{3} & \makebox[2em]{4}  &  \makebox[2em]{5} \\ \hline
         $D\Omega^d_O(B\mathrm{U(1)})$ & 0 & $\mathbb Z_2$ & 0 & $\mathbb Z_2^2$ & 0 & $\mathbb Z_2^4$ \\ \hline\hline
    \end{tabular}
    \caption{Cobordism groups $D\Omega_{O}^d(B\mathrm{U(1)})$.}
    \label{tbl:MO}
\end{table}

\begin{table}[t!]
    \centering
    \begin{tabular}{c*{6}{c}}\hline \hline
         $d$ & \makebox[2em]{0} & \makebox[2em]{1} & \makebox[2em]{2}  & \makebox[2em]{3} & \makebox[2em]{4}  &  \makebox[2em]{5} \\ \hline
         $D\Omega^d_O(B(\mathrm{U(1)}\times \mathbb Z^2))$ & 0 & $\mathbb Z_2$ & $\mathbb Z_2^2$ & $\mathbb Z_2\oplus \mathbb Z_2^2$ & $\mathbb Z_2^4$ &   $\mathbb Z_2^2 \oplus \mathbb Z_2^4$ \\ \hline\hline
    \end{tabular}
    \caption{Cobordism groups $D\Omega_{O}^d(B(
    \mathrm{U(1)}\times \mathbb Z^2))$.}
    \label{tbl:MO2}
\end{table}

\section{LSM-type anomaly as stacked half odd-integer spin}\label{app:top}

In this section, we give an interpretation of the LSM-type anomaly in view of the generalized cohomology classification.
For this purpose, we use the identification of the LSM-type anomaly in ($2+1$)-dimensional systems observed in the main text with the surface anomaly emerging in a weak SPT phase in ($3+1$)-dimensional systems protected by the $\mathrm{U}(1)\times\mathbb{Z}_2^T\times\mathbb{Z}^2$ symmetry. 
In particular, we focus on the physical understanding of the subgroup $\mathbb{Z}_2\subset D\Omega^{3}_O\left(B\mathrm{U}(1)\right)$ which describes the ingappability observed in the main text.

\subsection{Topological-field-theoretical classifications}

The LSM-type anomaly with which we are concerned corresponds to the surface anomaly of the weak SPT phase in $3+1$ dimensions classified by $D\Omega^{d=5}_O(B(\mathrm{U(1)}\times \mathbb Z^2))$ in table~\ref{tbl:MO2}.
Here, the latter is further reduced to
\begin{align}
    D\Omega^5_O(B(\mathrm{U(1)}\times \mathbb Z^2))
    &= D\Omega^3_O(B\mathrm{U(1)}) \oplus D\Omega^5_O(B\mathrm{U(1)})
    \notag \\
    &= \mathbb Z_2^2 \oplus \mathbb Z_2^4,
    \label{eq:fuC}
\end{align}
where the subgroup $D\Omega^{5}_O\left(B\mathrm{U}(1)\right)\simeq \mathbb{Z}_2^4$ represents the ($3+1$)-dimensional SPT phases protected only by U(1) and time-reversal symmetries independently of the translation. 
In such phases, the surface theory of the SPT phase  is free from the anomaly originating in the translational symmetry.
In other words, the surface theory cannot have a trivial ground state even in the absence of translational symmetry. 
This situation does not fit into the lattice model of our interest.
In fact, the lattice model can be gapped trivially by a translational symmetry-violating perturbation.

\begin{table}[t!]
	\begin{tabular}{c|*{7}{c}}
		5 & $\mathbb{Z}_2^2$ & 0  & $\mathbb{Z}_2^2$ & 0  & $\mathbb{Z}_2^2$  & 0  &  $\mathbb{Z}_2^2$    \\
		4 & 0 & 0 & 0 & 0 & 0  & 0  & 0  \\
		3 & $\mathbb{Z}_2$ & 0  & $\mathbb{Z}_2$ & 0  & $\mathbb{Z}_2$     &  0    &   $\mathbb{Z}_2$    \\
		2 & 0 & 0 & 0 & 0 & 0  & 0  & 0  \\
		1 & $\mathbb{Z}_2$ & 0  & $\mathbb{Z}_2$ & 0  & $\mathbb{Z}_2$     &  0    &  $\mathbb{Z}_2$    \\
		0 & 0 & 0 & 0 & 0 & 0  & 0  &  0  \\\hline
		\diagbox[dir=NE]{$q$}{$p$}
			&\makebox[3em]{0}&\makebox[3em]{1}&\makebox[3em]{2}&\makebox[3em]{3}
			&\makebox[3em]{4}&\makebox[3em]{5}&\makebox[3em]{6}\\
	\end{tabular}
	\caption{The $E_2$ page of the Atiya-Hirzebruch spectral sequence Eq.~(\ref{eq:E2pageAHss}). $p+q$ corresponds to $d$ in table.~\ref{tbl:MO}.}\label{tbl:E2pageAHss}
\end{table}

The remaining subgroup $D\Omega^{3}_O\left(B\mathrm{U}(1)\right)\simeq \mathbb{Z}_2^2$ in Eq.~(\ref{eq:fuC}) represents the ($3+1$)-dimensional weak SPT phases constructed by stacking ($1+1$)-dimensional SPT phases protected by U(1) and time-reversal symmetries. In such weak SPT phases, the translational symmetry plays the role of an obstacle that prohibits the system from becoming trivially gapped when being stacked.
To see the nature of the elements in $D\Omega^{3}_O\left(B\mathrm{U}(1)\right)\simeq \mathbb{Z}_2^2$, we consider the Atiya-Hirzebruch spectral sequence of the generalized cohomology theory:
\begin{align}
    D\Omega^{d}_O\left(B\mathrm{U}(1)\right) \Longleftarrow E_2^{p,q}= H^p\left(B\mathrm{U}(1);D\Omega^{q}_O(pt)\right). 
    \label{eq:E2pageAHss}
\end{align}
Here, $d$ on the left hand side corresponds to $p+q$.
The $E_2$ page is shown in table.~\ref{tbl:E2pageAHss}.
Note that $\oplus_{p+q=d} E_\infty^{p,q}$ is equal as a set to $D\Omega^d_O(B\mathrm{U(1)})$.
In addition, we can see in Eq.~\eqref{eq:E2pageAHss} that $E_2^{p,q} = E_\infty^{p,q}$ holds true for $p+q\le 5$.
Tables~\ref{tbl:E2pageAHss} and \ref{tbl:MO} turn out to be consistent with each other.

We are interested in the $d=5$ case in Eq.~\eqref{eq:E2pageAHss} corresponding to the ($3+1$)-dimensional weak SPT phase.
One can construct the latter by stacking ($1+1$)-dimensional SPT phases represented by the $d=3$ case in Eq.~\eqref{eq:E2pageAHss}, that is,
$D\Omega^3_O(B\mathrm{U(1)}) \simeq \mathbb Z_2^2$.
The $E_\infty$ page of the spectral sequence leads to
\begin{align}
    &D\Omega^3_O(B\mathrm{U(1)})
    \notag \\
    &\simeq H^0 \bigl(B\mathrm{U(1)}; D\Omega^3_O(pt)\bigr) \oplus H^2\bigl( B\mathrm{U(1)}; D\Omega^1_O(pt) \bigr)
    \notag \\
    &\simeq D\Omega^3_O(pt) \oplus H^2\bigl( B\mathrm{U(1)}; D\Omega^1_O(pt) \bigr)
    \label{eq:do3e2}
    \\
    &\simeq \mathbb Z_2 \oplus \mathbb Z_2.
\end{align}
In Eq.~\eqref{eq:do3e2}, $D\Omega^3_O(pt) \simeq \mathbb Z_2$ represents the ($1+1$)-dimensional SPT phase protected only by the time-reversal symmetry.
The effective action on its ($0+1$)-dimensional surface is deduced in the following.
The universal property of the Anderson dual gives
\begin{align}
    D\Omega^{d}_O(pt)\simeq
	\mathrm{Ext}^1_{\mathbb{Z}}(\Omega_{d-1}^O(pt),\mathbb{Z})\oplus \mathrm{Hom}_{\mathbb{Z}}(\Omega_{d}^O(pt),\mathbb{Z}),
\end{align}
where $\Omega_d^O(pt)$ is the $d$-dimensional unoriented bordism group. From the definition of the Ext functor, the short exact sequence $0\rightarrow \mathbb{Z} \rightarrow \mathbb{R} \rightarrow \mathbb{R}/\mathbb{Z}\rightarrow 0$ induces the following long exact sequence:
\begin{widetext}
\begin{align}
	0&\rightarrow \mathrm{Hom}_{\mathbb{Z}}(\Omega_{d}^O(pt),\mathbb{Z})\rightarrow \mathrm{Hom}_{\mathbb{Z}}(\Omega_{d}^O(pt),\mathbb{R})\rightarrow \mathrm{Hom}_{\mathbb{Z}}(\Omega_{d}^O(pt),\mathbb{R}/\mathbb{Z})\nonumber\\
	&\rightarrow \mathrm{Ext}_{\mathbb{Z}}^1(\Omega_{d}^O(pt),\mathbb{Z})\rightarrow \mathrm{Ext}_{\mathbb{Z}}^1(\Omega_{d}^O(pt),\mathbb{R})\rightarrow \mathrm{Ext}_{\mathbb{Z}}^1(\Omega_{d}^O(pt),\mathbb{R}/\mathbb{Z})\nonumber\\
	&\rightarrow\cdots.
\end{align}
\end{widetext}
We note that $\Omega_d^O(pt)$ is a two-torsion group because for a bordism class $[M]\in \Omega_d^O(pt)$, $2[M]=\partial[M\times I]=0$. This fact leads to $\mathrm{Hom}_{\mathbb{Z}}(\Omega_{d}^O(pt),\mathbb{R})=0$ and $\mathrm{Ext}_{\mathbb{Z}}^1(\Omega_{d}^O(pt),\mathbb{R})=0$ in the above long exact sequence. We thus have
\begin{align}
    \mathrm{Hom}_{\mathbb{Z}}(\Omega_{d}^O(pt),\mathbb{R}/\mathbb{Z})
	\simeq \mathrm{Ext}_{\mathbb{Z}}^1(\Omega_{d}^O(pt),\mathbb{Z}),
\end{align}
which relates the Pontryagin dual \cite{Kapustin:2014tfa, Kapustin:2014dxa} and the Anderson dual of the unoriented bordism groups. Consequently, we have
\begin{align}
    D\Omega^{3}_O(pt)\simeq
	\mathrm{Hom}_{\mathbb{Z}}(\Omega_{2}^O(pt),\mathbb{R}/\mathbb{Z}).
\end{align}
Note that the unoriented bordism groups are characterized by the Stiefel-Whitney numbers.
Therefore, the generator of $D\Omega^{3}_O(pt)$ finally turns out to be $\exp\left(i\pi \int  w_1\smile w_1\right)$, as already specified in Ref.~\cite{Kap14}. On an oriented spacetime $M$, $w_1=\delta \eta$ for a cochain $\eta\in C^0(M;\mathbb{Z}_2)$ and the topological action becomes $\exp\left[i\pi \int_M  \delta( \eta\delta\eta)\right]$. 
If $M$ has a surface $\partial M$, the surface-effective action becomes
\begin{eqnarray}
	\exp\left(i\pi \int_{\partial M} \eta\delta\eta \right).
\end{eqnarray}
This effective action is not invariant under the gauge transformation $\eta\rightarrow \eta +\theta$ ($w_1\rightarrow w_1+\delta \theta$). 
Now suppose $\partial M$ is a triangle whose vertices are labeled by the numbers 0, 1 and 2, and suppose $\eta(0)=1, \eta(1)=\eta(2)=0$, which signifies that the time-reversal operation acts on the system twice along the time direction. In this situation, the partition function $\exp\left(i\pi \int_{\partial M} \eta\delta\eta \right)$ takes on the value of $-1$.
Therefore, the anomaly represents the Kramers doublet in (0+1)d.

Let us consider the other part $H^2(B\mathrm{U(1)}; D\Omega^1_O(pt))$ of Eq.~\eqref{eq:do3e2}.
\begin{align}
    H^2\bigl(B\mathrm{U(1)}; D\Omega^1_O(pt) \bigr) \simeq \mathbb Z_2
\end{align}
which is generated by the mod-$2$ reduction of the first Chern class $c_1$ as already specified in Ref.~\cite{Kap14}.
The topological action $\exp(i\pi \int_M c_1)$ on a spacetime $M$ without the monopole is given by $\exp(i\pi \int_M \delta a)$, where $\delta a= c_1$.
When $M$ has a surface $\partial M$, the surface-effective action becomes
\begin{align}
    \exp\biggl(i\pi \int_{\partial M} a \biggr).
\end{align}
Here, the anomaly emerges as the noninvariance of the ($0+1$)-dimensional surface theory under the large gauge transformation of $a \to a+\theta$ ($c_1 \to c_1 + \delta \theta$) with $\theta \in C^1(B\mathrm{U(1)}; \mathbb Z)\otimes \mathbb{Z}_2$.
Therefore, the anomaly is characterized by a half-odd-integer $\mathrm{U(1)}$ charge~\cite{Kap14}.

\subsection{Interpretation of the anomaly}

These topological-field-theoretical characterizations of the anomalies can be understood intuitively.
Let us recall that our lattice model is composed of half-odd-integer spins on each site.
The half-odd-integer spin is the Kramers doublet and, at the same time, has a half-odd-integer $\mathrm{U(1)}$ charge.
We are thus led to the fact that the relevant 't Hooft anomaly in our quantum spin systems is the element
\begin{align}
    (1,1) \in D\Omega^3_O(pt) \oplus H^2\bigl(B\mathrm{U(1)}; D\Omega^1_O(pt) \bigr),
\end{align}
which is merely the surface anomaly of the spin-1 chain in the Haldane phase.
We thus reach the following reasonable interpretation of the LSM-type anomaly.
Each site is equipped with a half-odd-integer spin whose eigenstate is doubly degenerate.
The degeneracy cannot be lifted because the translation symmetry forbids the stacking of such a spin with nearby spins.
This interpretation is consistent with the lattice homotopy argument~\cite{po_lattice_homotopy}.

\subsection{Spin-1 Haldane phase as the $\mathrm{U(1)}\times \mathbb Z_2^T$ SPT phase}\label{app:haldane}

In the above discussion, the spin-$1$ Haldane phase is identified with the element $(1,1) \in D\Omega^3_O(pt) \oplus H^2\bigl(B\mathrm{U(1)}; D\Omega^1_O(pt) \bigr)$ as ($1+1$)-dimensional U(1)$\times\mathbb{Z}_2^T$ SPT. It is worth noting that the spin-$1$ Haldane phase has the topological action of $\exp(i\pi\int c_1)$ in addition to $\exp(i\pi\int w_1^2)$. To see this, we check that the spin-1 Haldane phase exhibits a nontrivial response to a monopole insertion into the ($1+1$)-dimensional spacetime, since the topological action $\exp(i\pi\int c_1)$ counts the number of monopoles. 
To realize the monopole insertion in the operator formalism, we employ the twist operator $U$ of the original LSM theorem in (1+1)-dimensions~\cite{lsm, nakamura_twist}. 
\begin{align}
	U:=\exp\biggl(i\frac{2\pi}{L}\sum_{n=1}^{L}n(S-S^z_n)\biggr).
\end{align} 
In the following, we argue that the operator $U$ acting on quantum spin chains inserts a monopole to the (1+1)-dimensional spacetime. More precisely, we argue that the ground-state expectation value of $U$ is the partition function (i. e. the generating functional of response functions) $Z[A]$ in the presence of the external U(1)-gauge field $A$ created by a monopole.

The nature of $U$ is clarified in the continuum limit of vanishing lattice spacing $a\to0$ with fixed system size $l_x=La=\mathrm{const}$, where $S-S^z_n$ is regarded as the U(1)-charge density $n_c(x=na)$. 
Consequently, the operator $U$ can be regarded as the minimal coupling term between the charge density $n_c(t,x)$ and the U(1)-gauge field $\mathbf{A}(t,x)$:
\begin{align}
	U=\exp\left(i\int_0^{l_x} \int_{-T/2}^{T/2} dxdt \,  \mathbf{A}\cdot \mathbf{j}\right),
\end{align}
where 
\begin{align}
	A^0(t,x)=\frac{2\pi}{l_x}\delta(t) x,\; A^1(t,x)=0,\label{eq:u1gauge}\\
	j^0(t,x)=n_c(t,x),\;j^1(t,x)=0.
\end{align}
The partition function in the presence of the external gauge field is obtained by the expectation value of the minimal coupling term, namely,
\begin{align}
	Z[A]=\left\langle \exp\left(i\int dxdt\,  \mathbf{A}\cdot \mathbf{j}\right)\right\rangle=\langle \mathrm{GS}|U|\mathrm{GS}\rangle.
\end{align}
Now we are ready to show that the external gauge field~\eqref{eq:u1gauge}  represents a monopole. 
From Eq.~\eqref{eq:u1gauge}, the field strength $F$ of the external gauge field becomes
\begin{align}
	F=dA=\frac{2\pi}{l_x}\delta(t) dx\wedge dt,
\end{align}
and the first Chern number (i.e. the number of monopole) becomes
\begin{align}
	\int\frac{F}{2\pi}=\int_0^{l_x} \int_{-T/2}^{T/2} dxdt\, \frac{1}{l_x}\delta(t)=1.
\end{align}
We can thus conclude that the external gauge field Eq.~(\ref{eq:u1gauge}) is indeed created by a monopole.

We have shown that the ground-state expectation value of $U$ is the partition function $Z[A]$ in the presence of a monopole. If the system is in an SPT phase $(1,1) \in D\Omega^3_O(pt) \oplus H^2\bigl(B\mathrm{U(1)}; D\Omega^1_O(pt) \bigr)$, the partition function $Z[A]$ contains a nontrivial phase factor of the topological action $\exp\left(i\pi\int c_1\right)$, and thus the ground-state expectation value of $U$ must contain a nontrivial phase factor of $e^{i\pi}=-1$. Indeed, in the spin-1 Haldane phase, the ground-state expectation value of the operator $U$ contains the nontrivial phase factor, which means that the twist operator $U$ is an order parameter of the Haldane phase~\cite{nakamura_twist}.

\bibliography{ref.bib}

\end{document}